\documentclass{elsart}
\usepackage{graphicx,amssymb}
\usepackage[centertags]{amsmath}
\usepackage{amssymb,amsfonts}
\usepackage{latexsym}

\newcommand{\be}{\begin{equation}} \newcommand{\ee}{\end{equation}}
\newcommand{\bea}{\begin{eqnarray}} \newcommand{\eea}{\end{eqnarray}}
\newcommand{\bse}{\begin{subequations}} \newcommand{\ese}{\end{subequations}}
\newcommand{\n}{\nonumber}

\begin{document}
\begin{frontmatter}

\title{First integrals for charged perfect fluid distributions}
\author{M.C Kweyama\thanksref{now},}
\thanks[now]{Permanent address: Department of Mathematical Sciences, 
Mangosuthu University of Technology, P. O. Box 12363, Jacobs, 4026, Durban, South Africa.}
\author{S.D. Maharaj\corauthref{cor},} 
\corauth[cor]{Corresponding author.}
\ead{Maharaj@ukzn.ac.za}
\author{K.S. Govinder}
\address{Astrophysics and Cosmology Research Unit,
School of Mathematical Sciences, University of KwaZulu-Natal,
Private Bag X54001, Durban 4000, South Africa.}

\begin{abstract}
We study the evolution of shear-free spherically symmetric charged fluids
in general relativity. We find
a new parametric class of solutions to the Einstein-Maxwell system of field equations.
Our charged results are a generalisation of earlier treatments for neutral relativistic fluids.
 We regain the first integrals found previously for uncharged matter as a special case.
 In addition an explicit first integral is found which is necessarily charged.\end{abstract}

\end{frontmatter}

\section{Introduction}
Solutions of the Einstein-Maxwell system of equations are important
in relativistic astrophysics as they may be used to describe charged
compact objects with strong gravitational fields such as dense
neutron stars.  Several recent treatments, including the works of
Ivanov \cite{Ivanov} and Sharma \textit{et al} \cite{Sharma},
demonstrate that the presence of the electromagnetic field affects
the values of redshifts, luminosities and maximum mass of a compact
relativistic star. The electromagnetic field cannot be ignored when
considering the gravitational evolution of stars composed of quark
matter as pointed out by Mak and Harko \cite{Mak} and Komathiraj and
Maharaj \cite{Komathiraj}.
 Therefore exact models describing the formation  and evolution of charged stellar objects, within
 the context of full general relativity, are necessary.  Electromagnetic fields play a role in gravitational
 collapse, the formation of naked singularities, and the collapse of charged shells of matter onto
 existing black holes (as indicated by Lasky and Lun \cite{Lasky, Lun}).  Significant electric fields are also present
 in phases of intense dynamical activity, in collapsing configurations, with time scales of the
 order of the hydrostatic time scale for which the usual stable equilibrium configurations assumptions
  are not reliable (as shown in the treatments of Di Prisco \textit{et al} \cite{Di} and Herrera \textit{et al} \cite{Herrera}).
  It is interesting to note
  that Maxwell's equations play a role in several other scenarios, including the evolution of cosmological
  models in higher dimensions.  De Felice and Ringeval \cite{De} considered braneworld models, exhibiting
  Poincare symmetry in extra-dimensions, which admit wormhole configurations.

  Spherical symmetry and a shear-free matter distribution are simplifying assumptions usually
made when seeking exact solutions to the Einstein field equations with neutral matter.
The field equations may then be reduced to a single partial differential equation. What is interesting about this equation is that it can be treated
as an ordinary differential equation. A general
class of solutions was first found by Kustaanheimo and Qvist \cite{Kustaanheimo}.  Comprehensive treatments
of the uncharged case are provided by Srivastava \cite{Srivastava} and Sussman \cite{Suss89}.  The generalisation to
include the electromagnetic field is easily  performed and is described by the Einstein-Maxwell system.  The field
equations are again reducible to a single partial differential equation, now containing a term corresponding
 to charge.  A review of known charged solutions, admitting a Friedmann limit, is contained in the treatment
 of Krasinski \cite{Krasinski}.  A detailed investigation of the mathematical and physical features of the Einstein-Maxwell
  system has been performed by Srivastava \cite{SrivaDC} and Sussman \cite{Suss88a, Suss88b} respectively.

The objective of this paper is to investigate the integrability
properties of the governing partial differential equation that
contains a term corresponding to charge, for shear-free fluids. This
investigation is performed using an elementary approach suggested by
Srivastava \cite{Srivastava}.  In Section 2, we reduce the
Einstein-Maxwell field equations, generalising the transformation
due to Faulkes \cite{Faulkes}, to a single nonlinear second order
partial differential equation that governs the behaviour of charged
fluids. As in the uncharged case, this equation can be treated as an
ordinary differential equation.
 In Section 3, we derive a first integral of the governing
equation by generalising the technique of Srivastava \cite{Srivastava} first used for uncharged fluids.  This first integral
is subject to two integrability conditions expressed as nonlinear integral equations.  We transform the integrability
 conditions, in Section 4, into a new system of differential equations which can be integrated in terms of quadratures.
  In Section 5 we comprehensively investigate the nature of the factors of the quartic arising in the quadrature.
   Finally, in Section 6, we discuss the results obtained and comment on some of the physical aspects.
   The approach of Khalique \textit{et al} \cite{k} using a group classification may be helpful in providing further insights
   in future work.

\section{Field equations}
We consider the shear-free motion of a spherically symmetric perfect
fluid in the presence of the electromagnetic field.  We choose a
coordinate system $x^{i}=(t,r,\theta,\phi)$ which is both comoving
and isotropic.  In this coordinate system the metric can be written
as
\begin{equation}
ds^{2}=-e^{2\nu(t,r)}dt^{2}+e^{2\lambda(t,r)}\left[dr^{2}+r^{2}\left(d\theta^{2}
+\sin^{2}\theta d\phi^{2}\right)\right]\label{eqna01}
\end{equation}
where $\nu$ and $\lambda$ are the gravitational potentials.  We are investigating the
general case of a self-gravitating fluid in the presence of the electromagnetic field without
 placing arbitrary restrictions on the potentials.  For this model the Einstein equations
 are supplemented with Maxwell equations.  The Einstein field equations for a charged
 perfect fluid can be written as the system
\begin{subequations}
\label{eqn2}
\begin{eqnarray}
\rho&=&3\frac{\lambda_{t}^{2}}{e^{2\nu}}-\frac{1}{e^{2\lambda}}\left(2\lambda_{rr}
+\lambda_{r}^{2}+\frac{4\lambda_{r}}{r}\right)-\frac{E^{2}}{r^{4}e^{4\lambda}} \label{eqn01a}\\
p&=&\frac{1}{e^{2\nu}}\left(-3\lambda_{t}^{2}-2\lambda_{tt}+2\nu_{t}\lambda_{t}\right)
+\frac{1}{e^{2\lambda}}\left(\lambda_{r}^{2}+2\nu_{r}\lambda_{r}+\frac{2\nu_{r}}{r}
+\frac{2\lambda_{r}}{r}\right) \nonumber \\
&& +\frac{E^{2}}{r^{4}e^{4\lambda}} \label{eqn01b}\\
p&=&\frac{1}{e^{2\nu}}\left(-3\lambda_{t}^{2}-2\lambda_{tt}+2\nu_{t}\lambda_{t}\right)
+\frac{1}{e^{2\lambda}}\left(\nu_{rr}+\nu_{r}^{2}+\frac{\nu_{r}}{r}+\frac{\lambda_{r}}{r} 
 +\lambda_{rr}\right) \nonumber \\
 && -\frac{E^{2}}{r^{4}e^{4\lambda}} \label{eqn01c}\\
0&=&\nu_{r}\lambda_{t}-\lambda_{tr} \label{eqn01d}
\end{eqnarray}
\end{subequations}
Maxwell's equations yield
\begin{eqnarray}
E&=&r^{2}e^{\lambda-\nu} \Phi_{r}, \qquad E_{r} =\sigma r^{2}e^{3\lambda}\label{eqn02}
\end{eqnarray}
In the above $\rho$ is the energy density and $p$ is the isotropic pressure
 which are measured relative to the four-velocity $u^{a}=(e^{-\nu},0,0,0)$. Subscripts refer to partial derivatives with respect to that variable.
 The quantity $E=E(r)$ is an arbitrary constant of integration and $\sigma$
 is the proper charge density of the fluid.  We interpret $E$ as the total charge
 contained within the sphere of radius $r$ centred around the origin of the coordinate
 system.  Note that $\Phi_{r}=F_{10}$ is the only nonzero component of the electromagnetic
 field tensor $F_{ab}=\phi_{b;a}-\phi_{a;b}$ where $\phi_{a}=\left(\Phi(t,r),0,0,0\right)$.
 The Einstein-Maxwell system (\ref{eqn2})-(\ref{eqn02}) is a coupled system of equations in
 the variables $\rho$, $p$, $E$, $\sigma$, $\nu$ and $\lambda$.

The system of partial differential equations (\ref{eqn2}) can be simplified to produce an
underlying nonlinear second order equation.  Equation (\ref{eqn01d}) can be written as
\[
\nu_{r}=\left(\ln\lambda_{t}\right)_{r}
\]
Then (\ref{eqn01b}) and (\ref{eqn01c}) imply
\[
\left[e^{\lambda}\left(\lambda_{rr}-\lambda_{r}^{2}-\frac{\lambda_{r}}{r}\right)
+\frac{2E^{2}e^{-\lambda}}{r^{4}}\right]_{t}=0 \]
and the potential $\nu$ has been eliminated.  The Einstein field equations (\ref{eqn2})
can therefore be written in the equivalent form
\begin{subequations}
\label{eqn09}
\begin{eqnarray}
\rho&=&3e^{2h}-e^{-2\lambda}\left(2\lambda_{rr}+\lambda_{r}^{2}+\frac{4\lambda_{r}}{r}\right)
-\frac{E^{2}}{r^{4}e^{4\lambda}}  \label{eqn09a}\\
p&=&\frac{1}{\lambda_{t}e^{3\lambda}}\left[e^{\lambda}\left(\lambda_{r}^{2}
+\frac{2\lambda_{r}}{r}\right)-e^{3\lambda+2h}-\frac{E^{2}}{r^{4}e^{\lambda}}\right]_{t} \label{eqn09b}\\
e^{\nu}&=&\lambda_{t}e^{-h} \label{eqn09c}\\
e^{\lambda}\left(\lambda_{rr}-\lambda_{r}^{2}-\frac{\lambda_{r}}{r}\right)&=&
-\tilde{F}-\frac{2E^{2}}{r^{4}e^{\lambda}} \label{eqn09d}
\end{eqnarray}
\end{subequations}
for a  charged relativistic fluid.  In the above $h=h(t)$ and $\tilde{F}=\tilde{F}(r)$
are arbitrary constants of integration.  Equation (\ref{eqn09d}) is the
condition of pressure isotropy generalised to include the electric field.
 To find an exact solution of the field equations, we need to specify the
 functions $h$, $\tilde{F}$ and $E$ and solve equation (\ref{eqn09d})
 for $\lambda$.  We can then compute the quantities $\rho$ and $p$ from
 (\ref{eqn09a}) and (\ref{eqn09b}), and $\sigma$ follows from (\ref{eqn02}).

It is possible to write (\ref{eqn09d}) in a simpler form by eliminating
the exponential factor $e^{\lambda}$.  We use the transformation, first
introduced by Faulkes \cite{Faulkes} for neutral fluids, which has the adapted form
\[
x=r^{2}, \quad
y=e^{-\lambda},\quad
f(x)=\frac{\tilde{F}}{4r^{2}},\quad
g(x)=\frac{E^{2}}{2r^{6}}
\]
Then (\ref{eqn09d}) becomes
\begin{equation}
y_{xx}=f(x)y^{2}+g(x)y^{3} \label{eqn013}
\end{equation}
which is the fundamental equation governing the behaviour of a shear-free charged fluid.
Observe that (\ref{eqn013}) is a nonlinear partial differential equation since $y=y(t,x)$.
When $g=0$ then $y_{xx}=f(x)y^{2}$ for a neutral fluid which has been
studied by Maharaj \textit{et al }\cite{Maharaj} and others.

\section{A charged integral}

It would appear that we need to specify the functions $f(x)$ and $g(x)$ to
integrate (\ref{eqn013}).  However it is possible \textit{ab inito }to generate
a first integral without choosing $f(x)$ and $g(x)$ if we generalise a technique
 first suggested by Srivastava \cite{Srivastava}, and extended by Maharaj \textit{et al }\cite{Maharaj}.
 The first integral generated is subject to a system of integral equations in $f(x)$
 and $g(x)$ which can be rewritten as differential equations.

Rather than choose $f(x)$ and $g(x)$ we seek general conditions that reduce the order
of (\ref{eqn013}) to produce a first order differential equation. We can formally
integrate (\ref{eqn013}) to obtain
\begin{eqnarray}
y_{x}&=&\int f(x)y^{2}dx+\int g(x)y^{3}dx \nonumber\\
&=&f_{I}y^{2}+g_{I}y^{3}-2\int f_{I}yy_{x}dx-3\int g_{I}y^{2}y_{x}dx \label{eqna6}
\end{eqnarray}
For convenience we have used the notation
\begin{eqnarray*}
\int f(x) dx&=&f_{I}, \qquad \int g(x) dx =g_{I}
\end{eqnarray*}
Integrating $f_{I}yy_{x}$ by parts and utilising (\ref{eqn013}) gives the result
\begin{equation}
\int f_Iyy_{x}dx=f_{II}yy_{x}-\int f_{II}y_{x}^{2}dx-\int ff_{II}y^{3}dx
-\int gf_{II}y^{4}dx \label{eqna7}
\end{equation}
We substitute (\ref{eqna7}) in (\ref{eqna6}) to obtain
\begin{eqnarray}
y_{x}&=&f_{I}y^{2}+g_{I}y^{3}-2f_{II}yy_{x}+2\int f_{II}y_{x}^{2}dx
\nonumber\\
& & + 2\int ff_{II}y^{3}dx+2\int gf_{II}y^{4}dx-3\int g_{I}y^{2}y_{x}dx \label{eqna8}.
\end{eqnarray}
We continue this process and evaluate the integrals of $f_{II}y_{x}^{2}$,
$ff_{II}y^{3}$ and $gf_{II}y^{4}$ in (\ref{eqna8}) using integration by parts.
Eventually we arrive at the expression
\begin{eqnarray}
y_{x}&=&f_{I}y^{2}+g_{I}y^{3}-2f_{II}y y_{x}+2f_{III}y_{x}^{2}
+2\left(ff_{II}\right)_{I}y^{3}+2\left(gf_{II}\right)_{I}y^{4}\nonumber\\
& &-\frac{2}{3}\int \left\{\left[2ff_{III}+
3\left(ff_{II}\right)_{I}+\frac{3}{2}g_{I}\right]\left(\frac{dy^{3}}{dx}\right)\right\}dx\nonumber\\
& &-\int \left\{\left[gf_{III}
+2\left(gf_{II}\right)_{I}\right]\left(\frac{dy^{4}}{dx}\right)\right\}dx \label{eqna9}
\end{eqnarray}
For a meaningful result the integrals on the right hand side of (\ref{eqna9}) must be eliminated.

We note that these integrals can be determined
if $2ff_{III}+3\left(ff_{II}\right)_{I}+\frac{3}{2}g_{I}$
 and $gf_{III}+2\left(gf_{II}\right)_{I}$ are constants.
 This observation yields the following result
\begin{eqnarray}
\tau_{0}(t)&=&-y_{x}+f_{I}y^{2}+g_{I}y^{3}-2f_{II}y y_{x}
+2f_{III}y_{x}^{2}+2[\left(ff_{II}\right)_{I}-\frac{1}{3}K_{0}]y^{3} \nonumber \\
&& +[2\left(gf_{II}\right)_{I}-K_{1}]y^{4}\label{eqna10}
\end{eqnarray}
subject to the integrability conditions
\begin{subequations}
\label{eqna11}
\begin{eqnarray}
K_{0}&=&2ff_{III}+3\left(ff_{II}\right)_{I}+\frac{3}{2}g_{I} \label{eqna11a}\\
K_{1}&=&gf_{III}+2\left(gf_{II}\right)_{I} \label{eqna11b}
\end{eqnarray}
\end{subequations}
where $K_{0}$ and $K_{1}$ are constants, and the quantity
$\tau_{0}(t)$ is an arbitrary function of integration.
 We have therefore established that a first integral of
 the field equation (\ref{eqn013}) is given by (\ref{eqna10})
 subject to conditions (\ref{eqna11}) which are integral equations.
 On the surface it appears that the functions $f(x)$ and $g(x)$ are free.  
 However equations (\ref{eqna11a}) and (\ref{eqna11b}) effectively determine the 
 forms of the functions $f(x)$ and $g(x)$; they are constrained by the integrability conditions (\ref{eqna11}).

\section{Integrability conditions}

It is not easy to solve the nonlinear integral equations (\ref{eqna11}).  However we
can transform these equations into an equivalent system comprising a first order
and a fourth order ordinary differential equation which are more convenient to work with.

We let
\[
f_{III}={\cal F} \label{eqna13}
\]
so that $f_{II}={\cal F}_{x}$, $f_{I}={\cal F}_{xx}$ and $f={\cal F}_{xxx}$.
Then it is possible to rewrite (\ref{eqna11b}) as
\begin{equation}
\left(g{\cal F}\right)_{x}+2g{\cal F}_{x}=0 \label{eqna15}
\end{equation}
Note that the integral equation (\ref{eqna11b}) has been transformed to a first
order differential equation in ${\cal F}$.  Equation (\ref{eqna15}) is integrable and we obtain
\begin{equation}
g={\cal K}_{0}{\cal F}^{-3} \label{eqna17}
\end{equation}
where ${\cal F}={\cal F}(x)$ and ${\cal K}_{0}$ is an arbitrary constant.

Similarly we can eliminate $g$ in (\ref{eqna11a}), with the
help of (\ref{eqna17}), to get the result
\begin{equation}
{\cal F}{\cal F}_{xxxx}+\frac{5}{2}{\cal F}_{x}{\cal F}_{xxx}
=-\frac{3}{4}{\cal K}_{0}{\cal F}^{-3} \label{eqna20}
\end{equation}
Therefore the integral equation (\ref{eqna11a}) has been transformed
to a fourth order differential equation in ${\cal F}$.
Equation (\ref{eqna20}) can be integrated repeatedly to yield
\begin{eqnarray}
{\cal F}^{-1}&=&{\cal K}_{4}+{\cal K}_{3}\int {\cal F}^{-3/2}dx+{\cal K}_{2}\left(\int {\cal F}^{-3/2}dx\right)^{2}-\frac{1}{6}{\cal K}_{1}\left(\int {\cal F}^{-3/2}dx\right)^{3}\nonumber\\
& &+\frac{1}{32}{\cal K}_{0}\left(\int {\cal F}^{-3/2}dx\right)^{4}\label{eqna28}
\end{eqnarray}
where the ${\cal K}_{i}$ are arbitrary constants.

We can rewrite (\ref{eqna28}) in a simpler form if we let
\begin{equation}
u=\int {\cal F}^{-3/2}dx \label{uf}
\end{equation}
so that
\[
u_{x}=\left({\cal F}^{-1}\right)^{3/2}
\]
Then we can write (\ref{eqna28}) as
\begin{equation}
u_{x}=\left({\cal K}_{4}+{\cal K}_{3}u+{\cal K}_{2}u^{2}-\frac{1}{6}{\cal K}_{1}u^{3}+\frac{1}{32}{\cal K}_{0}u^{4}\right)^{3/2}\label{eqna31}
\end{equation}
which is a first order equation in $u$.  The equivalent integral representation is
\begin{equation}
x-x_{0}=\int\frac{du}{\left({\cal K}_{4}+{\cal K}_{3}u+{\cal K}_{2}u^{2}-(1/6){\cal K}_{1}u^{3}+(1/32){\cal K}_{0}u^{4}\right)^{3/2}} \label{eqna32}
\end{equation}
where $x_{0}$ is a constant.  The quadrature (\ref{eqna32}) can be evaluated in terms of elliptic integrals.  We can summarise our result as follows: \textit{the first integral (\ref{eqna10}), with $g={\cal K}_{0}{\cal F}^{-3}$, $f={\cal F}_{xxx}$ and ${\cal F}$ given by (\ref{eqna32}) via (\ref{uf}), represents a particular class of solutions of (\ref{eqn013}).}

To obtain solutions in closed form, satisfying the integrability conditions (\ref{eqna11}), we need to evaluate the integral (\ref{eqna32}).  Particular solutions in terms of elementary functions are admitted.  In general the solution will be given in terms of special functions.  We can express the solutions to (\ref{eqna11}) in the parametric form as follows
\begin{subequations}
\label{eqna33}
\begin{eqnarray}
f(x)&=&{\cal F}_{xxx}\label{eqna33a}\\
g(x)&=&{\cal K}_{0}{\cal F}^{-3}\label{eqna33b}\\
u_{x}&=&{\cal F}^{-3/2}=[G'(u)]^{-1}\label{eqna33c}\\
x-x_{0}&=&G(u)\label{eqna33d}
\end{eqnarray}
\end{subequations}
where we have set
\begin{equation}
G(u)=\int\frac{du}{\left({\cal K}_{4}+{\cal K}_{3}u+{\cal K}_{2}u^{2}-(1/6){\cal K}_{1}u^{3}+(1/32){\cal K}_{0}u^{4}\right)^{3/2}}\label{eqna034}
\end{equation}

If we set $g=0$ (which forces ${\cal K}_{0}=0$) then the charge vanishes and the system (\ref{eqna33}) becomes
\begin{subequations}
\label{eqna34}
\begin{eqnarray}
f(x)&=&{\cal F}_{xxx}\label{eqna34a}\\
u_{x}&=&{\cal F}^{-3/2}=[G'(u)]^{-1}\label{eqna34c}\\
x-x_{0}&=&G(u)\label{eqna34d}
\end{eqnarray}
\end{subequations}
where
\begin{equation}
G(u)=\int\frac{du}{\left({\cal K}_{4}+{\cal K}_{3}u+{\cal K}_{2}u^{2}-(1/6){\cal K}_{1}u^{3}\right)^{3/2}}\label{eqna35},
\end{equation}
This corresponds to the results found by Maharaj \textit{et al} \cite{Maharaj} for a neutral shear-free gravitating fluid.
Thus their first integral  is contained in our class of charged models (\ref{eqna33})-(\ref{eqna034}).

\section{Particular solutions}

Nine cases arise from the solution (\ref{eqna33})--(\ref{eqna034}) depending on the nature of the factors of the polynomial ${\cal K}_{4}+{\cal K}_{3}u+{\cal K}_{2}u^{2}-(1/6){\cal K}_{1}u^{3}+(1/32){\cal K}_{0}u^{4}$.

\subsection{Case I: One order-four linear factor}
If ${\cal K}_{4}+{\cal K}_{3}u+{\cal K}_{2}u^{2}-(1/6){\cal K}_{1}u^{3}+(1/32){\cal K}_{0}u^{4}$ has one repeated linear factor then we have \[{\cal K}_{4}+{\cal K}_{3}u+{\cal K}_{2}u^{2}-(1/6){\cal K}_{1}u^{3}+(1/32){\cal K}_{0}u^{4}=(a+bu)^{4}, b\neq0\]  We evaluate the integral in (\ref{eqna034}) to obtain
\begin{subequations}
\label{eqn035}
\begin{eqnarray}
G(u)&=&-\frac{1}{5b}(a+bu)^{-5}\label{eqna35a}\\
f(x)&=&\frac{24}{75}(5b)^{4/5}(x-x_{0})^{-11/5}\label{eqna35b}\\
g(x)&=&{\cal K}_{0}(5b)^{-12/5}(x-x_{0})^{-12/5}\label{eqna35c}
\end{eqnarray}
\end{subequations}
In this case it is possible to invert the integral (\ref{eqna32}) and then write $u=u(x)$.  The first integral (\ref{eqna10}) has the form
\begin{eqnarray}
\tau_{0}(t)&=&-y_{x}-\frac{4}{15}(5b)^{4/5}\left(x-x_{0}\right)^{-6/5}y^{2}-\frac{5}{7}{\cal K}_{0}(5b)^{-12/5}\left(x-x_{0}\right)^{-7/5}y^{3}\nonumber\\
& &-\frac{8}{3}(5b)^{4/5}\left(x-x_{0}\right)^{-1/5}yy_{x}+\frac{10}{3}(5b)^{4/5}\left(x-x_{0}\right)^{4/5}y_{x}^{2}\nonumber\\
& &-2\left[\frac{3856}{10815}(5b)^{8/5}\left(x-x_{0}\right)^{-7/5}-\frac{15}{14}{\cal K}_{0}(5b)\left(x-x_{0}\right)^{-7/5}\right]y^{3}\nonumber\\
& &-\frac{5}{3}{\cal K}_{0}(5b)^{-8/5}\left(x-x_{0}\right)^{-8/5}y^{4}\label{eqna135}
\end{eqnarray}
where we have used the functional forms in (\ref{eqn035}).  The first integral (\ref{eqna135}) corresponds to  a shear-free spherically symmetric charged fluid which does not have an uncharged  limit since ${\cal K}_{0}\neq 0$.  If ${\cal K}_{0}=0$ then the polynomial becomes cubic which is a contradiction.  The charged integral (\ref{eqna135}) ($E\neq 0, {\cal K}_{0}\neq 0, b\neq 0$) is a new solution to the Einstein-Maxwell field equations.

\subsection{Case II: One order-three  linear factor}
If ${\cal K}_{4}+{\cal K}_{3}u+{\cal K}_{2}u^{2}-(1/6){\cal K}_{1}u^{3}+(1/32){\cal K}_{0}u^{4}$ has two linear factors, one of which is not repeated, then we have \[{\cal K}_{4}+{\cal K}_{3}u+{\cal K}_{2}u^{2}-(1/6){\cal K}_{1}u^{3}+(1/32){\cal K}_{0}u^{4}=\left(a+bu\right)\left(u+c\right)^{3}\]  We use the computer package Mathematica \cite{Wolfram} to determine the integral in (\ref{eqna034}) to obtain
\begin{eqnarray}
G(u)&=&\frac{2\sqrt{(a+bu)(u+c)}}{35(a-bc)^{5}}\left[\frac{35b^{4}}{a+bu}+\frac{93b^{3}}{u+c}-\frac{29b^{2}(a-bc)}{(u+c)^{2}}
 \right.\nonumber\\
& & \left. +\frac{13b(a-bc)^{2}}{(u+c)^{3}}-\frac{5(a-bc)^{3}}{(u+c)^{4}}\right]\label{eqna35d}
\end{eqnarray}
expressed completely in terms of elementary functions.  In this case, if $g=0$, ${\cal K}_{0}=0$ and $b=0$, then (\ref{eqna35d}) becomes
\begin{equation}
G(u)=a^{-3/2}\left(-\frac{2}{7}\right)(u+c)^{-7/2}\label{eqna35e}
\end{equation}
and hence using (\ref{eqna34}) we find
\begin{equation}
f(x)=a^{2/7}\left(\frac{48}{343}\right)\left(-\frac{7}{2}\right)^{6/7}\left(x-x_{0}\right)^{-15/7}\label{eqna35f}
\end{equation}
Note that (\ref{eqna35f}) is related to the result obtained by Maharaj \textit{et al} \cite{Maharaj}.

Again setting $g=0$, $K_{1}=0$, in (\ref{eqna10}) we get
\[
\psi_{0}(t)=-y_{x}+f_{I}y^{2}+g_{I}y^{3}-2f_{II}y y_{x}+2f_{III}y_{x}^{2}+2[\left(ff_{II}\right)_{I}-\frac{1}{3}K_{0}]y^{3}
\]
which was the first integral for uncharged matter found by Maharaj \textit{et al} \cite{Maharaj}. 
 Also observe that if $g=0, K_{1}=0, f(x)=(ax+b)^{-15/7}$ then (\ref{eqna10}) yields
\begin{eqnarray}
\phi_{0}(t)&=&-6y_{x}-\frac{21}{4a}(ax+b)^{-8/7}y^{2}-\frac{3}{2}\left(\frac{7}{a}\right)^{2}(ax+b)^{-1/7}yy_{x}
\nonumber\\
& &+\frac{1}{4}\left(\frac{7}{a}\right)^{3}(ax+b)^{6/7}y_{x}^{2}-\frac{1}{6}\left(\frac{7}{a}\right)^{3}(ax+b)^{-9/7}y^{3}\label{eqna12b}
\end{eqnarray}
which was found by Srivastava \cite{Srivastava}.  Also with $g=0, K_{1}=0, f(x)=x^{-15/7}$ in (\ref{eqna10}) (or if we set $a=1, b=0$ in (\ref{eqna12b})) we have
\begin{eqnarray*}
\varphi_{0}(t)&=&-6y_{x}-\frac{21}{4}x^{-8/7}y^{2}-\frac{3}{2}\cdot7^{2}x^{-1/7}yy_{x}
+\frac{1}{4}\cdot7^{3}x^{6/7}y_{x}^{2} \\
&&-\frac{1}{6}\cdot7^{3}x^{-9/7}y^{3}
\end{eqnarray*}
which was established by Stephani \cite{Stephani}.  Therefore the first integral (\ref{eqna10}) is a charged generalisation of the particular Maharaj \textit{et al } \cite{Maharaj}, Srivastava \cite{Srivastava} and Stephani \cite{Stephani} neutral models.

\subsection{Case III: One order-two linear factor; one order-one quadratic factor}
If ${\cal K}_{4}+{\cal K}_{3}u+{\cal K}_{2}u^{2}-(1/6){\cal K}_{1}u^{3}+(1/32){\cal K}_{0}u^{4}$ 
has two factors, one linear and repeated and the other is irreducible 
to linear factors, then we have 
\begin{eqnarray*}
&&{\cal K}_{4}+{\cal K}_{3}u+{\cal K}_{2}u^{2}-(1/6){\cal K}_{1}u^{3}+(1/32){\cal K}_{0}u^{4} \\
&& =\left(a+bu+cu^{2}\right)\left(u+d\right)^{2}, b^{2}-4ac<0.
\end{eqnarray*}
 The function (\ref{eqna034}) is integrated to obtain
\begin{eqnarray}
&& G(u) = \nonumber \\
& &-\left\{\frac{1}{\left(a-bd+cd^{2}\right)u^{2}}+\frac{5(b-2cd)}{2(a-bd+cd^{2})u}\right.\nonumber\\
& &\left.-\frac{15(b-2cd)^{4}-62c(b-2cd)^{2}(a-bd+cd^{2})+24c^{2}(a-bd+cd^{2})^{2}}{2(a-bd+cd^{2})\left[4c(a-bd+cd^{2})-(b-2cd)^{2}\right]}\right.\nonumber\\
& &\left.-\frac{c(b-2cd)\left[15(b-2cd)^{2}-52c(a-bd+cd^{2})\right]u}{2(a-bd+cd^{2})\Delta}\right\}\times\nonumber\\
& &\frac{1}{2\sqrt{(a-bd+cd^{2})+(b-2cd)u+cu^{2}}} \nonumber \\
&& +\frac{15(b-2cd)^{2}-12c(a-bd+cd^{2})}{8(a-bd+cd^{2})^{3}} \times \nonumber\\
& &  \int\frac{du}{u\sqrt{a-bd+cd^{2}+(b-2cd)u+cu^{2}}}\label{eqna35g}
\end{eqnarray}
where $\Delta=4(a-bd+cd^{2})c-(b-2cd)^{2}$ and the integral on the right hand side can be expressed in terms of elementary functions.  The exact form of the integral depends on the signs of $a-bd+cd^{2}$ and $\Delta$ (see Gradshteyn and Ryzhik \cite{Gradshteyn}, equations 2.266 and 2.269.6).

\subsection{Case IV: One order-two linear factor; two order-one linear factors}
With one repeated and two non-repeated linear factors we have \[{\cal K}_{4}+{\cal K}_{3}u+{\cal K}_{2}u^{2}-(1/6){\cal K}_{1}u^{3}+(1/32){\cal K}_{0}u^{4}=(a+bu)(cu+d)(u+e)^{2}\]  In this case the expression for the integral in (\ref{eqna034}) can be evaluated with the help of the computer package Mathematica \cite{Wolfram}.  The resulting expression is expressible in terms of only elementary functions.  This expression is very lengthy and not illuminating, and is therefore not included in this work.

\subsection{Case V: Two order-two linear factors}
If ${\cal K}_{4}+{\cal K}_{3}u+{\cal K}_{2}u^{2}-(1/6){\cal K}_{1}u^{3}+(1/32){\cal K}_{0}u^{4}$ has two  linear factors each of which is repeated, then we have \[{\cal K}_{4}+{\cal K}_{3}u+{\cal K}_{2}u^{2}-(1/6){\cal K}_{1}u^{3}+(1/32){\cal K}_{0}u^{4}=(a+bu)^{2}(u+c)^{2}.\]  The integral in (\ref{eqna034}) may be easily determined so that
\begin{eqnarray}
G(u)&=&\left[6b^{2}\ln \frac{u+c}{a+bu}+\frac{3b^{2}(a-bc)}{a+bu}+\frac{b^{2}(a-bc)^{2}}{2(a+bu)^{2}}+\frac{3b(a-bc)}{u+c}\right.\nonumber\\
& &\left.-\frac{(a-bc)^{2}}{2(u+c)^{2}}\right]\frac{1}{(a-bc)^{5}}\label{eqna35h}
\end{eqnarray}
Thus for the case of two order-two linear factors the integral can be expressed completely in terms of elementary functions.

\subsection{Case VI: No repeated linear factors}
If ${\cal K}_{4}+{\cal K}_{3}u+{\cal K}_{2}u^{2}-(1/6){\cal K}_{1}u^{3}+(1/32){\cal K}_{0}u^{4}$ has no repeated linear factors, then we have \[{\cal K}_{4}+{\cal K}_{3}u+{\cal K}_{2}u^{2}-(1/6){\cal K}_{1}u^{3}+(1/32){\cal K}_{0}u^{4}=e(a+u)(b+u)(c+u)(d+u), e\neq0\]  In this case we obtain, in terms of elementary functions and elliptic integrals, the result \cite{Dieckmann}
\begin{eqnarray}
&& G(u) =\nonumber \\
&&\frac{2e^{-3/2}}{(a-b)\sqrt{(a+u)(b+u)(c+u)(d+u)}}\left[\frac{(a+u)(b+u)}{(b-c)(a-d)}\left[\frac{2}{(b-d)^{2}}\right.\right.\nonumber\\
& &\left.\left.\frac{1}{(b-d)(c-d)}+\frac{1}{(a-c)(c-d)+}\right]+\frac{b+u}{a-c}\left[\frac{2(d+u)}{(a-b)(a-d)^{2}}\right.\right.\nonumber\\
& &\left.\left.\frac{1}{(b-c)(b-d)}-\frac{1}{(a-d)(b-d)}\right]-\frac{1}{(b-c)(b-d)}\right]\nonumber\\
& &-\frac{4e^{-3/2}}{(a-b)\sqrt{b-d}}\left[\frac{1}{(a-d)^{2}(c-d)\sqrt{a-c}}+\frac{\sqrt{a-c}}{(a-b)(b-c)^{2}(b-d)}\right.\nonumber\\
& &\left.+\frac{a-b-c+d}{(c-d)^{2}(a-c)^{3/2}(b-c)}\right]E(\alpha,p) \nonumber \\
&& +\frac{2e^{-3/2}}{(a-c)^{3/2}(b-d)^{3/2}(b-c)(a-d)}\times\nonumber\\
& &\left[\frac{2(a+b-c-d)^{2}}{(b-c)(a-d)}+\frac{(a-b-c+d)^{2}}{(a-b)(c-d)}\right]F(\alpha,p), (0<d<c<b<a)\label{eqna35i}
\end{eqnarray}
where we have let \[\alpha=\arcsin\sqrt{\frac{(a-c)(d+u)}{(a-d)(c+u)}}, \qquad p=\frac{(b-c)(a-d)}{(a-c)(b-d)}\]  In (\ref{eqna35i}), $F(\alpha,p)$ is the elliptic integral of the first kind and $E(\alpha,p)$ is the elliptic integral of the second kind.  This result is similar to one of the results obtained by Maharaj \textit{et al} \cite{Maharaj}.  However their uncharged model is not regainable from the expression above as the polynomial here is necessarily quartic.

\subsection{Case VII: One order-two quadratic factor}
If ${\cal K}_{4}+{\cal K}_{3}u+{\cal K}_{2}u^{2}-(1/6){\cal K}_{1}u^{3}+(1/32){\cal K}_{0}u^{4}$ has one repeated quadratic irreducible factor, then we have \[{\cal K}_{4}+{\cal K}_{3}u+{\cal K}_{2}u^{2}-(1/6){\cal K}_{1}u^{3}+(1/32){\cal K}_{0}u^{4}=(a+bu+cu^{2})^{2}\]  In this case we obtain
\begin{eqnarray*}
G(u)&=&\frac{b+2cu}{4ac-b^{2}}\left[\frac{1}{2(a+bu+cu^{2})^{2}}+\frac{3c}{(4ac-b^{2})(a+bu+cu^{2})}\right]\\
& &+\frac{6c^{2}}{(4ac-b^{2)^{2}}}\int \frac{du}{a+bu+cu^{2}}
\end{eqnarray*}
which can be expressed in terms of only elementary functions.  The exact form of the integral depends on the sign of $4ac-b^{2}$ (see Gradshteyn and Ryzhik \cite{Gradshteyn}, equations 2.172 and 2.173.2).

\subsection{Case VIII: Two order-one quadratic factors}
With two non-repeated quadratic factors we have \[{\cal K}_{4}+{\cal K}_{3}u+{\cal K}_{2}u^{2}-(1/6){\cal K}_{1}u^{3}+(1/32){\cal K}_{0}u^{4}=(a+bu+cu^{2})(d+eu+u^{2})\]  In this case the expression for the integral in (\ref{eqna034}), using the computer package Mathematica \cite{Wolfram}, is obtainable but is not included in this work as it is very lengthy.  It may be expressed in terms of elementary functions and elliptic integrals.

\subsection{Case IX: One order-one cubic factor}
If ${\cal K}_{4}+{\cal K}_{3}u+{\cal K}_{2}u^{2}-(1/6){\cal K}_{1}u^{3}+(1/32){\cal K}_{0}u^{4}$ has one irreducible  cubic factor, then we have \[{\cal K}_{4}+{\cal K}_{3}u+{\cal K}_{2}u^{2}-(1/6){\cal K}_{1}u^{3}+(1/32){\cal K}_{0}u^{4}=(a+bu+cu^{2}+du^{3})(e+u)\]  The integral in (\ref{eqna034}) can again be found with the help of the computer package Mathematica \cite{Wolfram}.  It is given in terms of elementary functions, elliptic integrals and special functions.  However it is so lengthy that it is also not included in this work.

\section{Discussion}

In this paper we have modelled the behaviour of shear-free charged
fluids, and reduced the solution of the Einstein-Maxwell system of
field equations to a single nonlinear partial differential equation.
By treating this equation as an ordinary differential equation,
a first integral  was found using
elementary methods.   It is remarkable to note that the first
integral is obtainable without specifying the arbitrary functions
contained in the governing equation.  The first integral is subject
to a system of two integral equations which were replaced by a
system of two differential equations which can be integrated up to a
quadrature.  Consequently we have found a new class of parametric
solutions to the Einstein-Maxwell system for a charged gravitating
shear-free fluid.  The new solution is given by the parametric
equations (\ref{eqna33})-(\ref{eqna034}).

A detailed analysis of the factors of the quartic arising in the
quadrature was performed.  Two cases of interest arise.  Firstly, we
are in a position to explicitly invert the quadrature when there is
one repeated linear factor and explicitly write the first integral.
Then the model has to be necessarily charged.  We believe that this
is a new result.  Secondly, we can explicitly invert the quadrature
when there is one order-three linear factor.  This case contains that of
 vanishing charge and we regain the results of Maharaj 
 \textit{et al} \cite{Maharaj}, Srivastava \cite{Srivastava} and Stephani \cite{Stephani}.
In the remaining cases the functions $G(u)$ is a complicated combination
of elementary  functions or/and special functions. In these remaining cases
it is not possible, except maybe for special parameter values, to invert the 
integral and write expressions for $f(x)$ and $g(x)$ explicitly.

We make certain points, related to the mathematics, to clarify the approach followed in this paper.
Firstly, the Einstein-Maxwell system has been studied extensively in the past with
the objective of finding exact solutions. We have considered
earlier treatments and, in particular the comprehensive analyses of 
Krasinski \cite{Krasinski} and Stephani \cite{Stephani}, and have not found
any reference to the  first integrals established in this paper.
Secondly, we have generated the first integral (\ref{eqna10}) mathematically
following the approach of Maharaj  \textit{et al} \cite{Maharaj} for uncharged fluids.
We believe that this is an elegant approach and may be used in other investigations
for the gravitational field or other physical systems.
 Our new class of solutions may be useful in this
context, and could provide a deeper insight into the behaviour of
the gravitational field. A comprehensive mathematical analysis of
the integrability properties of (\ref{eqn013}) using the symmetry
properties of the equation may provide further solutions and
insights. For example the treatment of Halburd \cite{Halburd}, for the
uncharged shear-free case, established an equivalence with the
generalised Chazy equation and provided a new class of integrable
equations.

We now make certain comments, related to the physics, related to our results.
Firstly,  a natural question is whether the models generated here are physically meaningful.
It is difficult to perform a general qualitative study of the physical features of the models because of
the complexity of the functions involved.  However we have found explicit charged first integrals and
earlier uncharged first integrals are regained. Other parameter values may also
lead to acceptable models. These factors point to physical reasonability.
Secondly, the line element (\ref{eqna01}) is written in terms of isotropic and comoving coordinates; the spacetime is shear-free.
The introduction of charge, and assumptions made in the integration, do not affect (\ref{eqna01})
so that the spacetime remains shear-free. Thirdly, charged  shear-free models in the presence of heat flow are of
crucial importance in relativistic astrophysics \cite{Di, Herrera} and influence
the range of temperature profiles of the models discussed here.  In \cite{kk} it was shown that, for the most general spherically symmetric line element 
with acceleration, the causal transport equation reduces to
\be
\beta (qB)_{,t} T^{-\sigma} + A (q B) = - \alpha \frac{T^{3-\sigma}
(AT)_{,r}}{B} \label{nntemp1} \ee
where 
\be
\kappa = \gamma T^3 \tau_c  \,\,\,\,
\tau =
\left(\frac{\beta \gamma}{\alpha}\right) \tau_c = \beta T^{-\sigma}
\label{tau}
\ee
with $\gamma \geq0 , \alpha \geq 0, \beta \geq 0$ and $\sigma \geq 0$. Then it is possible to integrate this equation for 
general metric functions $A$ and $B$.
When $\beta=0$, all noncausal solutions of (\ref{nntemp1}) are given by
 \bea
(AT)^{4-\sigma}&=&\frac{\sigma-4}{\alpha} \int A^{4-\sigma} q B^2 {\rm d} r
+ F(t) \qquad \sigma\neq4 \label{nnnoncausg}\\
\ln (AT) &=& - \frac{1}{\alpha}\int qB^2{\rm d} r + F(t)
 \qquad \sigma=4 \label{nnnoncaus4}
 \eea
 where $F(t)$ is an arbitrary function of integration which is fixed by
 the expression for the temperature of the star at its surface $\Sigma$.
For causal solutions ($\beta\neq0$) two solutions are provided. 
In the case of constant mean collision time, {\it ie.} $\sigma=0$,
(\ref{nntemp1}) is simply integrated to yield
\be
(AT)^4 = - \frac{4}{\alpha} \left[\beta\int A^3 B (qB)_{,t}{\rm d} r + \int
A^4 q B^2 {\rm d} r\right] + F(t) \label{nncaus0} \ee
The only nonconstant mean collision time solution is given for $\sigma=4$:
\bea
(AT)^4 &=& -\frac{4 \beta}{\alpha}\exp\left(-\int\frac{4qB^2}{\alpha} {\rm
d} r\right) \int A^3 B (qB)_{,t} \exp\left(\int\frac{4qB^2}{\alpha} {\rm d}
r\right){\rm d} r \n \\
&&\mbox{}+ F(t) \exp\left(-\int\frac{4qB^2}{\alpha} {\rm d} r\right)
\label{nncaus4} \eea
Our results can be incorporated into the above framework with
$A = e^\nu$ and  $B = e^\lambda$  in the astrophysical context.

The main objective of this paper was to show that an earlier method to
obtain uncharged first integrals is extendible to the Einstein-Maxwell
system of equations. We have shown that this is possible and particular
charged first integrals have been explicitly obtained. It would be
beneficial to study the models generated in terms of the original
metric, at least in particular cases  for simple forms of $f(x)$ and $g(x)$,
and to evaluate the evolution of the charged relativistic fluid in terms
of the fluid and electromagnetic variables. Other quantities requiring 
attention are the spacetime structure, energy conditions, Riemann invariants
and causality. This is outside the scope of our present treatment and will be
considered in future work.

{\bf Acknowledgements}\\
MCK and KSG thank the National Research Foundation and the University of KwaZulu-Natal 
for financial support.  SDM acknowledges that this work is based on research supported by the 
South African Research Chair Initiative of the Department 
of Science and Technology and the National Research Foundation.

\end{document}